\documentclass[twocolumn]{pasj02} 
\usepackage{tabularx}
\usepackage{color}

\jyear{2024}
\Received{}%{yyyy/mm/dd}
\Accepted{}%{yyyy/mm/dd}
\begin{document} 

\title{ A growth of early superhump: Multi color observation of the WZ Sge star TCP J23580961+5502508 }

%%% begin:list of authors
% Do NOT capitalize all letters in "textsc".
\author{
 Ryosuke \textsc{Sazaki},\altaffilmark{1}\altemailmark \email{sazaki@astro.hiroshima-u.ac.jp} 
 Makoto \textsc{Uemura},\altaffilmark{2}\orcid{0000-0002-7375-7405}
 Tatsuya \textsc{Nakaoka},\altaffilmark{2}
 and 
 Ryo \textsc{Imazawa}\altaffilmark{1}\orcid{0000-0002-0643-7946}
}
\altaffiltext{1}{Department of Physics, Graduate School of Advanced Science and Engineering, Hiroshima University, 1-3-1 Kagamiyama, Higashi-Hiroshima, Hiroshima 739-8526, Japan}
\altaffiltext{2}{Hiroshima Astrophysical Science Center, Hiroshima University, Kagamiyama 1-3-1, Higashi-Hiroshima, Hiroshima 739-8526, Japan}

%\footnotetext[$\dag$]{Present address: ....}

%%% end:list of authors

%% !!! Select 3 to 5 words from PASJ's key words !!! 
%% List of Key Words: https://academic.oup.com/pasj/pages/Pasj_Keywords 
%% "\KeyWords{ }" always has to be placed before ``\maketitle'' 

\maketitle

\begin{abstract}
We report optical and near-infrared photometry of a WZ Sge-type dwarf nova, TCP J23580961+5502508, during its 2022 superoutburst, obtained using the 1.5-m Kanata telescope.
Our observation detected early superhumps on three consecutive nights which included the rising phase toward the peak of the outburst.
%%%The observed color of the early superhumps was reddest around the maximum on all three days.
The early superhumps exhibited a profile dominated by the primary maximum during the first two days, while a prominent secondary minimum appeared on the third day.
We reconstructed the structure of the accretion disk from multi-color light curves using the early superhump mapping.
The accretion disk has a prominent flaring structure on the leading side of the disk during the first two days.
An additional flaring structure emerged on the opposite side on the third day, forming a two-armed pattern that can be interpreted within the framework of the 2:1 resonance model.
The reconstructed disk structure in the first two days suggests the presence of an additional mechanism operating during the initial stage of early superhump development.

\end{abstract}

%\pagewiselinenumbers 

\section{Introduction}

Cataclysmic variables (CVs) are close binary systems consisting of a primary white dwarf and a secondary low-mass star filling the Roche lobe (for a review, see \cite{Warner1995}).
An accretion disk is formed around the white dwarf via Roche lobe overflow from the secondary in non-magnetic CVs.
Dwarf novae (DNe) are a subclass of CVs that exhibit recurrent outbursts with amplitudes of $2-8$ mag and recurrence times ranging from days to years.
It is widely accepted that the mechanism of DN outbursts can be explained by the disk instability model (for reviews, see \cite{Osaki1996}, \cite{Lasota2001}, \cite{Hameury2020}).

SU~UMa-type systems are a subclass of DNe characterized by long and bright outbursts, known as superoutbursts.
Between successive superoutbursts, SU~UMa stars also exhibit normal outbursts.
During superoutbursts, these systems show superhumps, which are single-peaked photometric modulations with periods a few percent longer than the orbital period.
When the accretion disk radius reaches the $3:1$ resonance radius, a tidal instability is triggered.
Superhumps are generally interpreted as arising from an eccentric, precessing accretion disk \citep{Whitehurst1988A, Osaki1989, Hirose&Osaki1990, Lubow1991a, Lubow1991b}.

WZ Sge-type stars form a subclass of SU~UMa-type systems, characterized by extremely small binary mass ratios ($q \equiv M_{\mathrm{secondary}}/M_{\mathrm{WD}})$. 
They exhibit an exceptionally long recurrence time (typically $\sim10$ years), as well as a large amplitude ($> 6$ mag) and a long duration ($\sim$ a month) of superoutburst while normal outbursts are observed extremely rarely or not at all.
A hallmark of WZ Sge-type systems is the presence of "early superhumps" --- double-peaked modulations observed during the earliest phase of outbursts. 

The period of early superhumps is nearly identical to the orbital period of the binary system \citep{Ishioka2002}.
Their amplitudes depend on the orbital inclination: while the amplitudes of most early superhumps are less than 0.05 mag, some systems with high inclination exhibit early superhumps exceeding 0.1 mag (\cite{Kato2015}; \cite{Kato2002A}).
Early superhumps also exhibit reddening in color indices near the brightness maxima of the light curves (\cite{Matsui2009}; \cite{Nakagawa2013}; \cite{Imada2018a}).
These features suggest that early superhumps are originated from geometric effects, specifically the rotation of a non-axisymmetric flaring disk (\cite{Kato2002A}; \cite{Osaki&Meyer2002}; \cite{Maehara2007}).
Theoretically, early superhumps are proposed to be triggered when the disk reaches the 2:1 resonance radius, where strong tidal torques from the secondary star act on the disk \citep{Osaki&Meyer2002}. 
%%%In contrast, ordinary superhumps, which are observed in SU UMa-type DNe, arise from the 3:1 resonance, which induces disk eccentricity and results in a precessing eccentric disk \citep{Whitehurst1988A, Osaki1989, Hirose&Osaki1990, Lubow1991a, Lubow1991b}. 
In contrast, ordinary superhumps arise from 3:1 resonance.
The 2:1 resonance is considered to suppress the growth of the 3:1 resonance, which explains why ordinary superhumps are observed only after the early superhump phase.

The 2:1 resonance secnario is a leading theoretical model for early superhumps.
However, its validity remains uncertain due to the lack of observations in the earliest stages of outbursts and the limited understanding of whether it can account for the observed color variations.
\citet{Uemura2012} proposed a tomography method, called the early superhump mapping, in which a height map of the disk is estimated from multi-band light curves.
However the number of studies with the early superhump mapping is limited, and the diversity of disk structures has not been thoroughly investigated. 
Although early superhumps change their light curve profiles over time, there have been no observational studies that have tracked these variations in multiple bands from the early stages of an outburst. 
As a result, the temporal evolution of the disk structure during the early superhump phase remains unexplored.

An outburst of TCP J23580961+5502508 (hereafter, TCP J2358) was discovered on September 30, 2022, as a new transient by Tadashi Kojima\footnote{http://www.cbat.eps.harvard.edu/unconf/followups/J23580961+5502508.html}.
The autonomous observation system "Smart Kanata" found the discovery report of TCP J2358 from "Transient Objects Confirmation Page (TOCP)\footnote{http://www.cbat.eps.harvard.edu/unconf/tocp.html}" and we began follow-up observations immediately after the report \citep{SmartK}.
It is likely that its counterpart in quiescence is Gaia DR3 1993876650919517440 with 
$\mathrm{G} = 20.6 \; \mathrm{mag}$ \citep{Gaia_dr3}.
\citet{Tampo2026} report the results of a detailed analysis of early and ordinary superhumps of TCP~J2358
using light curves obtained with \textit{Transiting Exoplanet Survey Satellite} (TESS).
According to \citet{Tampo2026}, the object has an early superhump period of $0.059371(3) \; \mathrm{d}$, a stage-A superhump period of $0.06105(2) \; \mathrm{d}$, and thereby an inferred mass ratio of $0.074(1)$.

In this paper, we present our optical and infrared simultaneous observation of TCP J2358 during its 2022 outburst.
These data allow us, for the first time, to investigate the evolution of the disk structure over a period of three consecutive nights, starting before the outburst maximum.
We describe the method of our observation and analysis in section \ref{sec:obs} and present a result in section \ref{sec:result}.
In section \ref{sec:esm}, we describe the overview of the early superhump mapping and application to our observation data.
We discuss the results of TCP J2358 in section \ref{sec:discuss}, and a summary of this paper is given in section \ref{sec:summary}.

\section{Observation and Reduction}\label{sec:obs}

Time-resolved CCD photometric observations of TCP J2358 were performed with the Hiroshima Optical and Near-InfraRed camera (HONIR), which enables simultaneous optical and near-infrared observations \citep{Akitaya2014honir}. 
This instrument is installed at the Cassegrain focus of the 1.5-m Kanata telescope at the Higashi-Hiroshima Observatory.
We used {\it V}- and {\it J}-band filters.
We took the images using a dithering mode to accurately subtract a foreground sky.
The observations were carried out over 15 days from 2022 September 30 to October 31.
The log of our photometric observations is given in Table \ref{tab:obslog}.

We applied bias or dark subtraction and sky background subtraction for data reduction.
The photometric magnitudes of TCP J2358 were measured using the standard aperture photometry method. 
We calculated the median of the zero-point magnitude using stars within the field of view listed in the Pan-STARRS1 catalog \citep{Pan-STARRS} for the optical band and in the Two Micron All Sky Survey (2MASS) catalog \citep{2MASS} for the infrared band.
We report magnitudes in the Vega magnitude system.

\begin{table*}[tb]
%    \captionsetup{justification=centering} % キャプションを中央寄せにする
    \caption{Observation log of TCP J2358 using Kanata}
    \centering
    \begin{tabularx}{\textwidth}{p{0.25\textwidth}XXXXX}
    \hline
    MJD & $T_{\mathrm{exp}, V} ^*$ & $T_{\mathrm{exp}, J}^*$ & $N^\dag$ & $V^\ddag$ & $J^\ddag$ \\
    \hline
    59852.5331 -- .8029 & 60 & 45 & 236 & $11.75(0.01)$ & $11.89(0.01)$ \\ % 59852.533142 to .802917
    59853.5199 -- .8251 & 60 & 45 & 252 & $12.11(0.01)$ & $12.07(0.01)$ \\ % 59853.519855 to .825116
    59854.4382 -- .8224 & 60 & 45 & 296 & $12.43(0.01)$ & $12.40(0.01)$ \\ % 59854.438189 to .822442
    59860.5640 -- .8212 & 60 & 45 & 212 & $13.58(0.02)$ & $13.28(0.04)$ \\ % 59860.564022 to .821209
    59862.4300 -- .6121 & 95 & 80 & 134 & $13.80(0.03)$ & $13.53(0.03)$ \\ % 59862.429954 to .612147
    59864.6510 -- .6673 & 75 & 60 & 11 & $13.94(0.01)$ & $13.74(0.01)$ \\ % 59864.651019 to .667309
    59865.7761 -- .7946 & 75 & 60 & 15 & $14.10(0.01)$ & $13.86(0.01)$ \\ % 59865.776134 to .794554
    59870.6834 -- .7019 & 75 & 60 & 11 & $14.45(0.01)$ & $14.35(0.07)$ \\ % 59870.683414 to .701921
    59871.7235 -- .7403 & --- & 60 & 10 & --- & $14.32(0.01)$ \\ % 59871.723455 to .740312
    59874.5487 -- .5545 & 95 & 80 & 5 & $14.83(0.02)$ & $14.7(0.1)$ \\ % 59874.548727 to .554514
    59875.6088 -- .6209 & 75 & 60 & 10 & $14.87(0.02)$ & $14.52(0.04)$ \\ % 59875.60875 to .620891
    59876.7616 -- .7732 & --- & 60 & 7 & --- & $15.5(0.3)$ \\ % 59876.761568 to .773154
    59877.6379 -- .6493 & 75 & 60 & 10 & $15.04(0.01)$ & $14.64(0.02)$ \\ % 59877.637876 to .649282
    59880.5179 -- .5248 & 75 & 60 & 5 & $17.33(0.04)$ & --- \\ % 59880.517934 to .524832
    59883.4761 -- .4817 & 90 & 75 & 5 & $17.96(0.08)$ & --- \\ % 59883.476065 to .481661
    \hline
    \multicolumn{4}{l}{* Exposure time in seconds.} \\
    \multicolumn{4}{l}{$\dag$ Number of images.} \\
    \multicolumn{4}{l}{$\ddag$ Average magnitudes in each day.}
    \end{tabularx}
    \\
    \vspace{1mm}
    \label{tab:obslog}
\end{table*}

\section{Results}\label{sec:result}

\begin{figure}[htbp]
    \begin{center}
        \includegraphics[width=80mm]{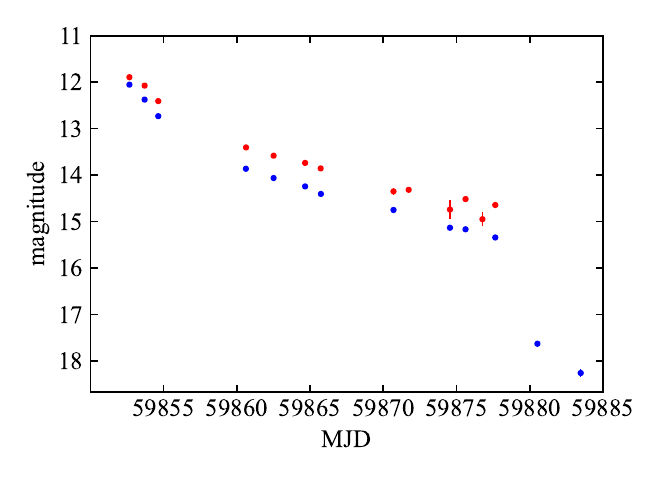}
    \end{center}
    \caption{Whole light curve of the outburst of TCP J2358 in 2022. The abscissa and ordinate denote the time in MJD and magnitude, respectively. 
    The blue and red circles represent the one-day averages of {\it V}- and {\it J}-band magnitudes, respectively. The {\it V}-band magnitudes are shifted by +0.3 mag for clarity.
    {Alt text: A scatter graph with error bars.}}
    \label{fig:whole_lc}
\end{figure}

Figure \ref{fig:whole_lc} shows the one-day averaged light curve of TCP J2358 during its 2022 outburst. 
Its peak magnitude was brighter than $11.8$ mag in the {\it V}-band on MJD 59852, which suggests that the amplitude of the outburst was $\sim 8.8$ mag. 
We calculated the mean fading rate using all data points, not one-day averaged ones.
The fading rate was $\sim0.34\; \mathrm{mag\;d^{-1}}$ in the {\it V}-band until MJD 59855 and then it decreased to $\sim0.09\; \mathrm{mag\;d^{-1}}$ from MJD 59860 to 59878. 
The outburst lasted at least 25 days and ended between MJD 59878 and 59880, with the magnitude suddenly dropping by $\sim 2.3 \;\mathrm{mag}$.
The decrease in the fading rate and the extended outburst duration are characteristic behaviors of superoutbursts in WZ Sge-type DNe (e.g., \cite{Nogami1997}).

\begin{figure}[htbp]
    \begin{center}
        \includegraphics[width=80mm]{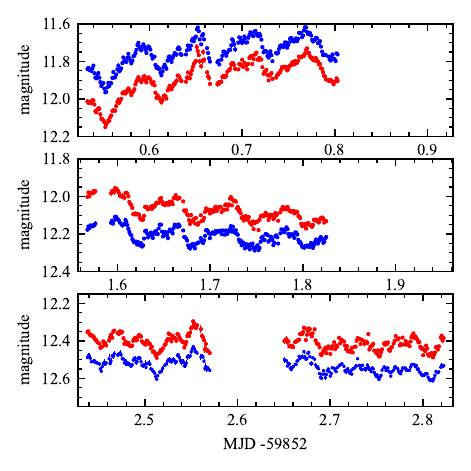}
    \end{center}
    \caption{
    The light curves of early superhumps observed in TCP J2358. 
    From top to bottom, the data were taken on MJD 59852 (Day 1), 59853 (Day 2), and 59854 (Day 3), respectively. 
    The blue and red circles represent the light curves in the {\it V}- and {\it J}-band, respectively. 
    For clarity, the {\it V}-band data on Days 2 and 3 are shifted by +0.1 mag.
    {Alt text: Three scatter graphs with error bars.}
    }
    \label{fig:eshlc}
\end{figure}

The observed light curves from MJD 59852 (Day 1) to MJD 59854 (Day 3)
are shown in Figure \ref{fig:eshlc}. 
Short-term periodic modulations are seen in all three days.
They have a doubly peaked profile, especially clearly on Days 2 and 3.
In conjunction with a late detection of ordinary superhumps (vsnet-alert 26980\footnote{http://ooruri.kusastro.kyoto-u.ac.jp/mailarchive/vsnet-alert/26980}), we conclude that the periodic modulations are early superhumps. 
In the top panel (Day 1), the object exhibited a gradually brightening trend with $-0.82$ and $-1.15\;\mathrm{mag}\;\mathrm{d^{-1}}$ in the {\it V}- and {\it J}-band light curves, respectively.
This suggests that the object was before the outburst maximum. 
After Day 2, the object started fading.
The fading rates were $0.31$ and $0.50\;\mathrm{mag}\;\mathrm{d^{-1}}$ in the {\it V}- and {\it J}-band on Day 2, and similarly, $0.14$ and $0.08 \;\mathrm{mag}\;\mathrm{d}^{-1}$ on Day 3.

\begin{figure}[htbp]
    \begin{center}
        \includegraphics[width=80mm]{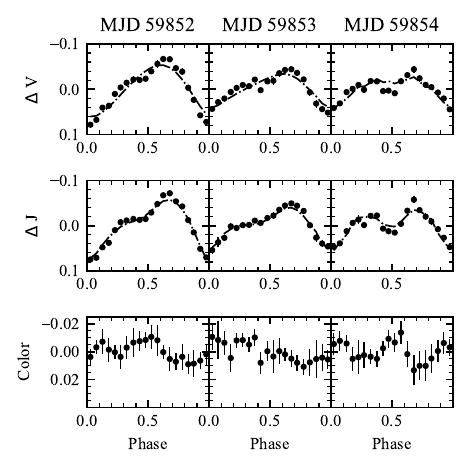}
    \end{center}
    \caption{
    Phase-averaged light curves of the early superhumps of TCP J2358 from Day 1 to Day 3. The top, middle, and bottom panels show {\it V}- and {\it J}-band light curves, as well as the color variation, respectively. The left, center, and right panels correspond to Days 1, 2, and 3. Filled circles in top and middle panels represent the observed light curve, while the dashed lines represent the model light curves.
    {Alt text: Nine-panel figure arranged in three rows and three columns. The top and middle rows plot scatter with error bars with dashed lines indicating the model. The bottom row plot scatter with error bars.}
    }
    \label{fig:binlc}
\end{figure}

The left, center and right panels of Figure \ref{fig:binlc} show the phase-averaged light curves and color for Days 1, 2, and 3, respectively. 
The global trend of each day's light curves was subtracted by assuming a quadratic component on Day 1 and a linear component on Days 2 and 3. 
The original light curves were folded using the early superhump period, $0.059371(3)$ d \citep{Tampo2026}, and averaged in 0.05-phase bins. 
The amplitude of the early superhump decreased from 0.16 on Day 1 to 0.10 on Day 2. 
Over time, the secondary minimum of the early superhump became more prominent. On Day 1, the hump exhibited a profile dominated by the primary maximum, particularly in the {\it V}-band, although a clear secondary peak appeared in the light curve by Day 3.

The $V - J$ color indices show a sign of becoming redder at or just after the primary maximum of the early superhumps.

\section{The early superhump mapping}\label{sec:esm}
We applied the early superhump mapping technique developed by \citet{Uemura2012} to J2358.
This tomography method reconstructs the disk height distribution, $h(r,\,\theta)$, of an accretion disk from multi-band light curves of early superhumps, under the assumption that they arise from rotational effects of a non-axisymmetric, optically thick disk.
For a given set of binary and disk hyperparameters (summarized in Table~\ref{tab:reconstruction_params}), the model flux $f_{\nu}(\phi)$ at frequency $\nu$ and orbital phase $\phi$ is calculated from $h(r,\,\theta)$.
The height distribution $h(r,\,\theta)$ is then optimized to reproduce the observed light curves using a regularized least-squares method. 
The regularization term includes a hyperparameter $w$ that controls the local smoothness of $h(r,\,\theta)$. 
In this study, we adopted $w = 2.0$, chosen as a compromise between suppressing overfitting and maintaining a good fit to the data.

For this analysis, we used the light curves shown in Figure \ref{fig:binlc}.
The values of the binary parameters used for the mapping are shown in Table~\ref{tab:reconstruction_params}.
We adopted the orbital period $P_{\mathrm{orb}} = P_{\mathrm{ESH}} = 0.059371 \; \mathrm{d}$
 and the mass ratio $q = 0.074$ from \citet{Tampo2026}.
%%%We confirmed that variations in $q$ had only a minor effect on the reconstructed structure of the accretion disk for TCP J2358, including its size, height, and the location of the flaring region.
The upper limit of the inclination angle of TCP J2358 is $70^{\circ}$ that is based on the lack of eclipses in its light curves.
The lower limit of the inclination angle can be given by the amplitude of the early superhump.
WZ Sge-type DNe that have early superhumps of high amplitude tend to have high inclination (\cite{Kato2015}).
The mean amplitude of the early superhump was $\sim 0.11$ mag.
This value is within the top $20 \%$ in amplitude among WZ Sge-type dwarf novae in which early superhumps have been observed \citep{Kato2015}.
This indicates that the object has a relatively high inclination.
%The amplitude of $0.15 \; \mathrm{mag}$ in TCP J2358 on Day 1 indicates that the object has a relatively high inclination. 
We thus set the inclination of TCP J2358 as $70^{\circ}$.
The innermost temperature of the accretion disk $T_\mathrm{in}$ was estimated from the observed average $V - J$ color of each day with the radius—temperature relation of the standard disk model, $T\propto r^{-3/4}$ \citep{ShakuraSunyaev1973}.
The mass of the white dwarf $M_{\mathrm{WD}}$ was adopted as $0.8\,M_\odot$, which is the typical value of white dwarf mass in DNe \citep{Zorotovic2011, Pala2022}.
We defined phase zero as the primary minimum of the early superhumps of {\it V}-band shown in Figure \ref{fig:eshlc}.
However, due to the lack of an accurate binary ephemeris, the orbital phase has an uncertainty of $\sim 0.1$, which corresponds to an uncertainty in the azimuthal direction of $\sim 36^\circ$ in the reconstructed disk structure.

\begin{table}[htbp]
    \caption{Model parameters for the early superhump mapping of TCP J2358.}
    %\vspace{2mm}
    \centering
    \begin{tabular}{ll}
    \hline
    Parameter  & Value  \\
    \hline
    $P_{\mathrm{orb}}(\mathrm{d})$  & $0.059371$  \\
    mass ratio $q$ & $0.074$ \\
    inclination angle  & $70^{\circ}$ \\
    $T_{\mathrm{in}}\;(\mathrm{K})$   & $88000$ (Day1), $69000$ (Day2), $70000$ (Day3)\\
    $R_{\mathrm{out}}$    & $0.6a$  \\
    $M_{\mathrm{WD}}$  & $0.8\, M_\odot$ \\
    \hline
    \end{tabular}
    \label{tab:reconstruction_params}
\end{table}

\begin{figure*}[tb]
    \begin{center}
        \includegraphics[width=160mm]{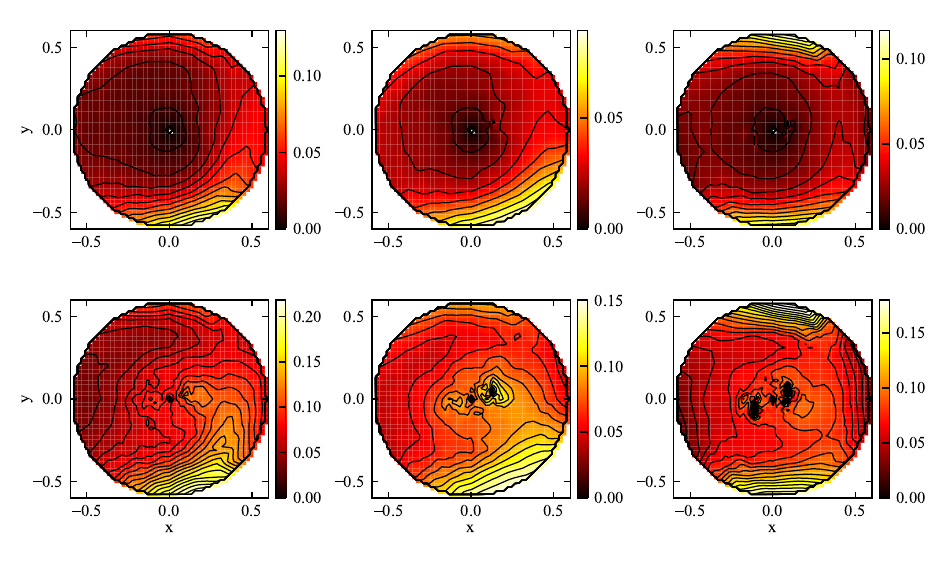}
    \end{center}
    \caption{
    Reconstructed height maps of the disk of TCP J2358 for the first three days. The upper and lower panels show the maps of height $h$ and $h/r$, respectively, where $r$ is the distance from the disk center. $h$, $x$, and $y$ are normalized by the binary separation. Both the contours and color-map represent the same map. The secondary star is located at $(x,\;y) = (1.0,\;0.0)$ and rotates in the direction $(x,\;y)\;=\;(1.0,\;1.0)$.
    {Alt text: Six-panel graph arranged in two rows and three columns. Color map with overlaid contour lines shows height of the accretion disk. Each panel includes a color bar on the right side.}
    }
    \label{fig:height-map}
\end{figure*}

The height maps of the reconstructed accretion disks are shown in Figure \ref{fig:height-map}.
The upper and lower panels show the estimated value of $h$%which is the disk height scaled by the binary separation $a$
, and $h/r$ where $r$ is the distance from the disk center, respectively.
The left, middle, and right panels represent the height maps on Day1, 2, and 3, respectively.
%%%The center of the secondary star is located at $(x,y)=(1.0,0.0)$, and the secondary rotates anti-clockwise.
The model light curves, calculated from the reconstructed disks, are shown as the dashed lines in Figure \ref{fig:binlc}.
These models successfully reproduced the overall trend of the observed light curves, although the local maxima and minima are smoothed out.
This is because the model favors a smooth disk structure \citep{Uemura2012}.
Allowing for more complex structures by adjusting the smoothness parameter results in overfitting, that is, a disk with unphysically fine structures.
Therefore, the present model provides the optimal balance between simplicity and accuracy.

%%%マップはy軸反転している

On Days 1 and 2, a flaring region is visible only in the lower part of Figure \ref{fig:height-map}, which corresponds to the leading side, i.e., the side facing the observer at orbital phase 0.0, at the outermost part of the disk.
On Day 3, an additional flaring region emerges in the upper part of the disk in Figure \ref{fig:height-map}, which corresponds to the trailing side. 
The highest points are located at $(x,\,y) = (+0.13, \, -0.58)\; (h=0.13)$, $(+0.47,\, -0.36) \; (h=0.08)$, and $(+0.13, \, +0.58)\; (h=0.12)$ on Day 1, 2, and 3, respectively.
These results indicate that the disk height kept $h/a \sim 0.1$ at maximum for the first three days. These heights are comparable to previous results of the early superhump mapping \citep{Uemura2012, Nakagawa2013, Tampo2022}.
Moreover, the size of the flaring regions gradually decreased with time.
In addition to these outermost flaring regions, an arm-like pattern extends from the lower side of the figure toward the inner part of the disk on Day 1 and 2, and a similar feature may be present on Day 3.
The right side of each panel, which corresponds to the portion of the disk closer to the secondary star, consistently appears to be more elevated than the opposite side in all three days.

The single-peaked structure in the light curves on Day 1 and 2 is responsible for the single prominent flaring region in the disk.
On the other hand, the appearance of a distinct double-peaked profile in the light curves on Day 3 corresponded to the emergence of two flaring regions in the disk.
Additionally, the elevated right side contributes to the formation of the secondary minimum, which is shallower than the primary minimum.

\section{Discussions} \label{sec:discuss}

In the reconstructed disks of other WZ Sge-type stars, such as V455 And \citep{Uemura2012}, OT J0120 \citep{Nakagawa2013}, PNV J00444033+4113068 \citep{Tampo2022}, two flaring regions were identified.
Our observation started before the outburst maximum.
Our results represent the first instance of a disk structure with a single prominent flaring region appearing only around the outburst maximum (Day 1 and 2).
The reconstructed disk on Day 3 exhibited a characteristic common to previous studies.

The weak secondary maximum in the light curves may be a characteristic feature of the early superhump in the early stage of WZ Sge superoutbursts.
A hint of this feature is also seen in WZ Sge itself \citep{Ishioka2002}.
During the 2001 outburst of WZ Sge, the primary maximum of early superhumps was more prominent than the secondary maximum, both before and around the outburst maximum.
This feature is also seen in V738 Hya and ASASSN-25ci reported in \citet{Tampo2026}.
To confirm this characteristic, it is necessary to observe the early superhumps of more WZ Sge-type DNe over several days around the outburst maximum.

\citet{Osaki&Meyer2002} proposed that early superhumps are caused by the $2{:}1$ resonance.
The $2{:}1$ resonance produces a two-armed spiral-shaped dissipation pattern in the accretion disk, which may have a strong impact on angular momentum transport \citep{Lin&Papaloizou1979}.
Accordingly, this scenario predicts the formation of two flaring regions at the outermost parts of the disk.
Both of these structures are visible in the reconstructed disk on Day 3 (Figure~\ref{fig:height-map}).
In contrast, on Days 1 and 2, the flaring region on the trailing side is significantly weaker than that on the leading side.
This asymmetry suggests that the disk structure in these earlier stages is not fully developed into the configuration expected from the $2{:}1$ resonance alone.
Interestingly, the prominent flaring structure on the leading side appears to resemble the bright arc obtained by Doppler tomography in the ordinary SU UMa-type dwarf nova HT Cas during its superoutburst \citep{Neustroev2020}.
In that study, the bright arc was interpreted as arising from irradiation of tidally thickened outer disk regions by the white dwarf and/or the inner disk.
These results may indicate that our current understanding of the early superhumps is incomplete, or that an additional mechanism---possibly related to another type of tidal deformation---contributes to the formation of asymmetric disk structures in the early stages of the outburst.

%\citet{Osaki&Meyer2002} proposed that early superhumps are caused by the 2:1 resonance.
%The $2:1$ resonance produces a two-armed spiral-shaped dissipation pattern in the disk, and it may have a strong effect on the angular momentum transfer with mass ratio $q = 0.1$ \citep{Lin&Papaloizou1979}.
%Hence, this scenario predicts two flaring regions at the outermost parts of the disk.
%Both of these regions can be observed in the reconstructed disk on Day 3, as shown in Figure \ref{fig:height-map}. 
%In contrast, the flaring region on the trailing side is much weaker than that in the leading side on Days 1 and 2.
%Our results suggest that, prior to the state explained by the 2:1 resonance scenario, there exists another state that cannot be explained by this mechanism alone and requires an additional mechanism.

\section{Summary}\label{sec:summary}
We performed simultaneous optical and near-infrared observations of a WZ Sge-type DN, TCP J2358, during its 2022 outburst. 
The outburst exhibited a large amplitude ($\sim 8.8$ mag) and clear early superhumps.
The early superhumps were detected for three consecutive nights starting from the day of outburst discovery, with the first detection occurring during the rising phase before the outburst peak. 
This marks the first time that time-resolved, multicolor photometry of early superhumps over multiple consecutive nights, including the pre-maximum stage, has been achieved for this object.
The early superhump profiles on Days 1 and 2 were dominated by the primary maximum, while a distinct secondary maximum appeared on Day 3.
%%%The color indices of the early superhumps became reddest around the primary maximum, consistent with the behavior observed in many other WZ Sge-type DNe.
We applied the early superhump mapping to reconstruct the structure of the accretion disk from the multi-wavelength light curves obtained on three days.
The reconstructed disk structures from Days 1 and 2 have a flaring region on the leading side.
In contrast, the reconstructed disk from the light curves of Day 3 exhibited an additional flaring structure on the trailing side, resembling the structure predicted by the 2:1 resonance model.

\bibliography{thesis}

@ARTICLE{Neustroev2020,
       author = {{Neustroev}, V.~V. and {Zharikov}, S.~V.},
        title = "{Voracious vortices in cataclysmic variables. II. Evidence for the expansion of accretion disc material beyond the Roche lobe of the accretor in HT Cassiopeia during its 2017 superoutburst}",
      journal = {\aap},
         year = 2020,
        month = oct,
       volume = {642},
          eid = {A100},
        pages = {A100},
          doi = {10.1051/0004-6361/201936597},
archivePrefix = {arXiv},
       eprint = {1908.10867},
 primaryClass = {astro-ph.SR},
       adsurl = {https://ui.adsabs.harvard.edu/abs/2020A&A...642A.100N}
}

@ARTICLE{Hameury2020,
       author = {{Hameury}, J.~M.},
        title = "{A review of the disc instability model for dwarf novae, soft X-ray transients and related objects}",
      journal = {Advances in Space Research},
         year = 2020,
        month = sep,
       volume = {66},
       number = {5},
        pages = {1004-1024},
          doi = {10.1016/j.asr.2019.10.022},
archivePrefix = {arXiv},
       eprint = {1910.01852},
 primaryClass = {astro-ph.SR},
       adsurl = {https://ui.adsabs.harvard.edu/abs/2020AdSpR..66.1004H}
}

@ARTICLE{Lasota2001,
       author = {{Lasota}, Jean-Pierre},
        title = "{The disc instability model of dwarf novae and low-mass X-ray binary transients}",
      journal = {New Astronomy Reviews},
         year = 2001,
        month = jun,
       volume = {45},
       number = {7},
        pages = {449-508},
          doi = {10.1016/S1387-6473(01)00112-9},
archivePrefix = {arXiv},
       eprint = {astro-ph/0102072},
 primaryClass = {astro-ph},
       adsurl = {https://ui.adsabs.harvard.edu/abs/2001NewAR..45..449L}
}

@ARTICLE{Zorotovic2011,
       author = {{Zorotovic}, M. and {Schreiber}, M.~R. and {G{\"a}nsicke}, B.~T.},
        title = "{Post common envelope binaries from SDSS. XI. The white dwarf mass distributions of CVs and pre-CVs}",
      journal = {\aap},
         year = 2011,
        month = dec,
       volume = {536},
          eid = {A42},
        pages = {A42},
          doi = {10.1051/0004-6361/201116626},
archivePrefix = {arXiv},
       eprint = {1108.4600},
 primaryClass = {astro-ph.SR},
       adsurl = {https://ui.adsabs.harvard.edu/abs/2011A&A...536A..42Z}
}

@ARTICLE{Pala2022,
       author = {{Pala}, A.~F. and {G{\"a}nsicke}, B.~T. and {Belloni}, D. and {Parsons}, S.~G. and {Marsh}, T.~R. and {Schreiber}, M.~R. and {Breedt}, E. and {Knigge}, C. and {Sion}, E.~M. and {Szkody}, P. and {Townsley}, D. and {Bildsten}, L. and {Boyd}, D. and {Cook}, M.~J. and {De Martino}, D. and {Godon}, P. and {Kafka}, S. and {Kouprianov}, V. and {Long}, K.~S. and {Monard}, B. and {Myers}, G. and {Nelson}, P. and {Nogami}, D. and {Oksanen}, A. and {Pickard}, R. and {Poyner}, G. and {Reichart}, D.~E. and {Rodriguez Perez}, D. and {Shears}, J. and {Stubbings}, R. and {Toloza}, O.},
        title = "{Constraining the evolution of cataclysmic variables via the masses and accretion rates of their underlying white dwarfs}",
      journal = {\mnras},
         year = 2022,
        month = mar,
       volume = {510},
       number = {4},
        pages = {6110-6132},
          doi = {10.1093/mnras/stab3449},
archivePrefix = {arXiv},
       eprint = {2111.13706},
 primaryClass = {astro-ph.SR},
       adsurl = {https://ui.adsabs.harvard.edu/abs/2022MNRAS.510.6110P}
}

@ARTICLE{Tampo2026,
       author = {{Tampo}, Y. and {Kojiguchi}, N. and {Isogai}, K. and {Nogami}, D. and {Itoh}, H. and {Hambsch}, F.-J. and {Matsumoto}, K. and {Matsumura}, R. and {Fujii}, D. and {Tordai}, T. and {Sano}, Y. and {Monard}, B. and {Dubovsky}, P.~A. and {Medulka}, T. and {Buckley}, D.~A.~H. and {Rawat}, N. and {Potter}, S.~B. and {van Dyk}, A. and {Groot}, P.~J. and {Woudt}, P. and {Kiyota}, S. and {Bolt}, G. and {Vanmunster}, T. and {Pietz}, J. and {Starr}, P. and {Shugarov}, S.~Y. and {Kasai}, K. and {Menzies}, K. and {Brincat}, S.~M. and {Pavlenko}, E.~P. and {Baklanov}, A. and {Ito}, J. and {Kato}, T.},
        title = "{TESS and ground-based observations of WZ Sge-type dwarf novae in outburst}",
      journal = {\mnras},
         year = 2026,
        month = jan,
       volume = {545},
       number = {2},
          eid = {staf1964},
        pages = {staf1964},
          doi = {10.1093/mnras/staf1964},
archivePrefix = {arXiv},
       eprint = {2511.04175},
 primaryClass = {astro-ph.SR},
       adsurl = {https://ui.adsabs.harvard.edu/abs/2026MNRAS.545f1964T}
}

@BOOK{Warner1995,
       author    = {Warner, Brian},
       title     = {Cataclysmic Variable Stars},
       year      = {1995},
       publisher = {Cambridge University Press},
       address   = {Cambridge},
       series    = {Cambridge Astrophysics Series},
       volume    = {28},
       adsurl    = {https://ui.adsabs.harvard.edu/abs/1995cvs..book.....W}
}

@ARTICLE{Imada2018a,
       author = {{Imada}, Akira and {Isogai}, Keisuke and {Araki}, Takahiro and {Tanada}, Shunsuke and {Yanagisawa}, Kenshi and {Kawai}, Nobuyuki},
        title = "{OAO/MITSuME photometry of dwarf novae. II. HV Virginis and OT J012059.6+325545}",
      journal = {\pasj},
         year = 2018,
        month = jan,
       volume = {70},
       number = {1},
          eid = {2},
        pages = {2},
          doi = {10.1093/pasj/psx142},
archivePrefix = {arXiv},
       eprint = {1711.06080},
 primaryClass = {astro-ph.SR},
       adsurl = {https://ui.adsabs.harvard.edu/abs/2018PASJ...70....2I}
}

@ARTICLE{Ishioka2002,
       author = {{Ishioka}, R. and {Uemura}, M. and {Matsumoto}, K. and {Ohashi}, H. and {Kato}, T. and {Masi}, G. and {Novak}, R. and {Pietz}, J. and {Martin}, B. and {Starkey}, D. and {Kiyota}, S. and {Oksanen}, A. and {Moilanen}, M. and {Cook}, L. and {Kral}, L. and {Hynek}, T. and {Kolasa}, M. and {Vanmunster}, T. and {Richmond}, M. and {Kern}, J. and {Davis}, S. and {Crabtree}, D. and {Beaulieu}, K. and {Davis}, T. and {Aggleton}, M. and {Gazeas}, K. and {Niarchos}, P. and {Yushchenko}, A. and {Mallia}, F. and {Fiaschi}, M. and {Good}, G.~A. and {Boyd}, D. and {Sano}, Y. and {Morikawa}, K. and {Moriyama}, M. and {Mennickent}, R. and {Arenas}, J. and {Ohshima}, T. and {Watanabe}, T.},
        title = "{First detection of the growing humps at the rapidly rising stage of dwarf novae AL Com and WZ Sge}",
      journal = {\aap},
        year = 2002,
        month = jan,
       volume = {381},
        pages = {L41-L44},
          doi = {10.1051/0004-6361:20011644},
archivePrefix = {arXiv},
       eprint = {astro-ph/0111432},
 primaryClass = {astro-ph},
       adsurl = {https://ui.adsabs.harvard.edu/abs/2002A&A...381L..41I}
}

@ARTICLE{Kato2002A,
       author = {{Kato}, Taichi},
        title = "{On the Origin of Early Superhumps in WZ Sge-Type Stars}",
      journal = {\pasj},
         year = 2002,
        month = apr,
       volume = {54},
        pages = {L11-L14},
          doi = {10.1093/pasj/54.2.L11},
archivePrefix = {arXiv},
       eprint = {astro-ph/0201233},
 primaryClass = {astro-ph},
       adsurl = {https://ui.adsabs.harvard.edu/abs/2002PASJ...54L..11K}
}

@ARTICLE{Kato2015,
       author = {{Kato}, Taichi},
        title = "{WZ Sge-type dwarf novae}",
      journal = {\pasj},
         year = 2015,
        month = dec,
       volume = {67},
       number = {6},
          eid = {108},
        pages = {108},
          doi = {10.1093/pasj/psv077},
archivePrefix = {arXiv},
       eprint = {1507.07659},
 primaryClass = {astro-ph.SR},
       adsurl = {https://ui.adsabs.harvard.edu/abs/2015PASJ...67..108K}
}

@ARTICLE{Lin&Papaloizou1979,
       author = {{Lin}, D.~N.~C. and {Papaloizou}, J.},
        title = "{Tidal torques on accretion discs in binary systems with extreme mass ratios.}",
      journal = {\mnras},
         year = 1979,
        month = mar,
       volume = {186},
        pages = {799-812},
          doi = {10.1093/mnras/186.4.799},
       adsurl = {https://ui.adsabs.harvard.edu/abs/1979MNRAS.186..799L}
}

@ARTICLE{Maehara2007,
       author = {{Maehara}, Hiroyuki and {Hachisu}, Izumi and {Nakajima}, Kazuhiro},
        title = "{Photometric Observation and Numerical Simulation of Early Superhumps in BC Ursae Majoris during the 2003 Superoutburst}",
      journal = {\pasj},
         year = 2007,
        month = feb,
       volume = {59},
        pages = {227-235},
          doi = {10.1093/pasj/59.1.227},
archivePrefix = {arXiv},
       eprint = {astro-ph/0611519},
 primaryClass = {astro-ph},
       adsurl = {https://ui.adsabs.harvard.edu/abs/2007PASJ...59..227M}
}

@ARTICLE{Matsui2009,
       author = {{Matsui}, Risako and {Uemura}, Makoto and {Arai}, Akira and {Sasada}, Ahito and {Ohsugi}, Takashi and {Yamashita}, Takuya and {Kawabata}, Koji and {Fukazawa}, Yasushi and {Mizuno}, Tsumefumi and {Katagiri}, Hideaki and {Takahashi}, Hiromitsu and {Sato}, Shuji and {Kino}, Masaru and {Yoshida}, Michitoshi and {Shimizu}, Yasuhiro and {Nagayama}, Shogo and {Yanagisawa}, Kenshi and {Toda}, Hiroyuki and {Okita}, Kiichi and {Kawai}, Nobuyuki},
        title = "{Optical and Near-Infrared Photometric Observation during the Superoutburst of the WZ Sge-Type Dwarf Nova, V455 Andromedae}",
      journal = {\pasj},
         year = 2009,
        month = oct,
       volume = {61},
        pages = {1081},
          doi = {10.1093/pasj/61.5.1081},
archivePrefix = {arXiv},
       eprint = {0908.4164},
 primaryClass = {astro-ph.SR},
       adsurl = {https://ui.adsabs.harvard.edu/abs/2009PASJ...61.1081M}
}

@ARTICLE{Nakagawa2013,
       author = {{Nakagawa}, Shinichi and {Noguchi}, Ryo and {Iino}, Eriko and {Ogura}, Kazuyuki and {Matsumoto}, Katsura and {Arai}, Akira and {Isogai}, Mizuki and {Uemura}, Makoto},
        title = "{Multicolor Photometry of an Outburst of a New WZ Sge-Type Dwarf Nova, OT J012059.6+325545}",
      journal = {\pasj},
         year = 2013,
        month = jun,
       volume = {65},
          eid = {70},
        pages = {70},
          doi = {10.1093/pasj/65.3.70},
archivePrefix = {arXiv},
       eprint = {1304.1855},
 primaryClass = {astro-ph.SR},
       adsurl = {https://ui.adsabs.harvard.edu/abs/2013PASJ...65...70N}
}

@ARTICLE{Nogami1997,
       author = {{Nogami}, Daisaku and {Kato}, Taichi and {Baba}, Hajime and {Matsumoto}, Katsura and {Arimoto}, Jun'ichi and {Tanabe}, Kazuhito and {Ishikawa}, Kaoru},
        title = "{The 1995 Superoutburst of the WZ Sge-type Dwarf Nova AL Comae Berenices}",
      journal = {\apj},
         year = 1997,
        month = dec,
       volume = {490},
       number = {2},
        pages = {840-846},
          doi = {10.1086/304881},
       adsurl = {https://ui.adsabs.harvard.edu/abs/1997ApJ...490..840N}
}

@ARTICLE{Osaki1996,
       author = {{Osaki}, Yoji},
        title = "{Dwarf-Nova Outbursts}",
      journal = {\pasp},
         year = 1996,
        month = jan,
       volume = {108},
        pages = {39},
          doi = {10.1086/133689},
       adsurl = {https://ui.adsabs.harvard.edu/abs/1996PASP..108...39O}
}

@ARTICLE{Osaki&Meyer2002,
       author = {{Osaki}, Y. and {Meyer}, F.},
        title = "{Early humps in WZ Sge stars}",
      journal = {\aap},
         year = 2002,
        month = feb,
       volume = {383},
        pages = {574-579},
          doi = {10.1051/0004-6361:20011744},
archivePrefix = {arXiv},
       eprint = {astro-ph/0112309},
 primaryClass = {astro-ph},
       adsurl = {https://ui.adsabs.harvard.edu/abs/2002A&A...383..574O}
}

@ARTICLE{ShakuraSunyaev1973,
       author = {{Shakura}, N.~I. and {Sunyaev}, R.~A.},
        title = "{Black holes in binary systems. Observational appearance.}",
      journal = {\aap},
         year = 1973,
        month = jan,
       volume = {24},
        pages = {337-355},
       adsurl = {https://ui.adsabs.harvard.edu/abs/1973A&A....24..337S}
}

@ARTICLE{Tampo2022,
       author = {{Tampo}, Yusuke and {Isogai}, Keisuke and {Kojiguchi}, Naoto and {Uemura}, Makoto and {Kato}, Taichi and {Tordai}, Tam{\'a}s and {Vanmunster}, Tonny and {Itoh}, Hiroshi and {Dubovsky}, Pavol A. and {Medulka}, Tom{\'a}{\v{s}} and {Sano}, Yasuo and {Hambsch}, Franz-Josef and {Taguchi}, Kenta and {Maehara}, Hiroyuki and {Ito}, Junpei and {Nogami}, Daisaku},
        title = "{PNV J00444033+4113068: Early superhumps with 0.7 mag amplitude and non-red color}",
      journal = {\pasj},
         year = 2022,
        month = dec,
       volume = {74},
       number = {6},
        pages = {1287-1294},
          doi = {10.1093/pasj/psac068},
archivePrefix = {arXiv},
       eprint = {2208.04251},
 primaryClass = {astro-ph.SR},
       adsurl = {https://ui.adsabs.harvard.edu/abs/2022PASJ...74.1287T}
}

@ARTICLE{Uemura2012,
       author = {{Uemura}, Makoto and {Kato}, Taichi and {Ohshima}, Tomohito and {Maehara}, Hiroyuki},
        title = "{Reconstruction of the Structure of Accretion Disks in Dwarf Novae from the Multi-Band Light Curves of Early Superhumps}",
      journal = {\pasj},
         year = 2012,
        month = oct,
       volume = {64},
          eid = {92},
        pages = {92},
          doi = {10.1093/pasj/64.5.92},
archivePrefix = {arXiv},
       eprint = {1203.1358},
 primaryClass = {astro-ph.GA},
       adsurl = {https://ui.adsabs.harvard.edu/abs/2012PASJ...64...92U}
}

@ARTICLE{Whitehurst1988A,
       author = {{Whitehurst}, Robert},
        title = "{Numerical simulations of accretion discs - I. Superhumps : a tidal phenomenon of accretion discs.}",
      journal = {\mnras},
         year = 1988,
        month = may,
       volume = {232},
        pages = {35-51},
          doi = {10.1093/mnras/232.1.35},
       adsurl = {https://ui.adsabs.harvard.edu/abs/1988MNRAS.232...35W}
}

@ARTICLE{Gaia_dr3,
       author = {{Gaia Collaboration} and {Vallenari}, A. and {Brown}, A.~G.~A. and {Prusti}, T. and {de Bruijne}, J.~H.~J. and {Arenou}, F. and {Babusiaux}, C. and {Biermann}, M. and {Creevey}, O.~L. and {Ducourant}, C. and {Evans}, D.~W. and {Eyer}, L. and {Guerra}, R. and {Hutton}, A. and {Jordi}, C. and {Klioner}, S.~A. and {Lammers}, U.~L. and {Lindegren}, L. and {Luri}, X. and {Mignard}, F. and {Panem}, C. and {Pourbaix}, D. and {Randich}, S. and {Sartoretti}, P. and {Soubiran}, C. and {Tanga}, P. and {Walton}, N.~A. and {Bailer-Jones}, C.~A.~L. and {Bastian}, U. and {Drimmel}, R. and {Jansen}, F. and {Katz}, D. and {Lattanzi}, M.~G. and {van Leeuwen}, F. and {Bakker}, J. and {Cacciari}, C. and {Casta{\~n}eda}, J. and {De Angeli}, F. and {Fabricius}, C. and {Fouesneau}, M. and {Fr{\'e}mat}, Y. and {Galluccio}, L. and {Guerrier}, A. and {Heiter}, U. and {Masana}, E. and {Messineo}, R. and {Mowlavi}, N. and {Nicolas}, C. and {Nienartowicz}, K. and {Pailler}, F. and {Panuzzo}, P. and {Riclet}, F. and {Roux}, W. and {Seabroke}, G.~M. and {Sordo}, R. and {Th{\'e}venin}, F. and {Gracia-Abril}, G. and {Portell}, J. and {Teyssier}, D. and {Altmann}, M. and {Andrae}, R. and {Audard}, M. and {Bellas-Velidis}, I. and {Benson}, K. and {Berthier}, J. and {Blomme}, R. and {Burgess}, P.~W. and {Busonero}, D. and {Busso}, G. and {C{\'a}novas}, H. and {Carry}, B. and {Cellino}, A. and {Cheek}, N. and {Clementini}, G. and {Damerdji}, Y. and {Davidson}, M. and {de Teodoro}, P. and {Nu{\~n}ez Campos}, M. and {Delchambre}, L. and {Dell'Oro}, A. and {Esquej}, P. and {Fern{\'a}ndez-Hern{\'a}ndez}, J. and {Fraile}, E. and {Garabato}, D. and {Garc{\'\i}a-Lario}, P. and {Gosset}, E. and {Haigron}, R. and {Halbwachs}, J. -L. and {Hambly}, N.~C. and {Harrison}, D.~L. and {Hern{\'a}ndez}, J. and {Hestroffer}, D. and {Hodgkin}, S.~T. and {Holl}, B. and {Jan{\ss}en}, K. and {Jevardat de Fombelle}, G. and {Jordan}, S. and {Krone-Martins}, A. and {Lanzafame}, A.~C. and {L{\"o}ffler}, W. and {Marchal}, O. and {Marrese}, P.~M. and {Moitinho}, A. and {Muinonen}, K. and {Osborne}, P. and {Pancino}, E. and {Pauwels}, T. and {Recio-Blanco}, A. and {Reyl{\'e}}, C. and {Riello}, M. and {Rimoldini}, L. and {Roegiers}, T. and {Rybizki}, J. and {Sarro}, L.~M. and {Siopis}, C. and {Smith}, M. and {Sozzetti}, A. and {Utrilla}, E. and {van Leeuwen}, M. and {Abbas}, U. and {{\'A}brah{\'a}m}, P. and {Abreu Aramburu}, A. and {Aerts}, C. and {Aguado}, J.~J. and {Ajaj}, M. and {Aldea-Montero}, F. and {Altavilla}, G. and {{\'A}lvarez}, M.~A. and {Alves}, J. and {Anders}, F. and {Anderson}, R.~I. and {Anglada Varela}, E. and {Antoja}, T. and {Baines}, D. and {Baker}, S.~G. and {Balaguer-N{\'u}{\~n}ez}, L. and {Balbinot}, E. and {Balog}, Z. and {Barache}, C. and {Barbato}, D. and {Barros}, M. and {Barstow}, M.~A. and {Bartolom{\'e}}, S. and {Bassilana}, J. -L. and {Bauchet}, N. and {Becciani}, U. and {Bellazzini}, M. and {Berihuete}, A. and {Bernet}, M. and {Bertone}, S. and {Bianchi}, L. and {Binnenfeld}, A. and {Blanco-Cuaresma}, S. and {Blazere}, A. and {Boch}, T. and {Bombrun}, A. and {Bossini}, D. and {Bouquillon}, S. and {Bragaglia}, A. and {Bramante}, L. and {Breedt}, E. and {Bressan}, A. and {Brouillet}, N. and {Brugaletta}, E. and {Bucciarelli}, B. and {Burlacu}, A. and {Butkevich}, A.~G. and {Buzzi}, R. and {Caffau}, E. and {Cancelliere}, R. and {Cantat-Gaudin}, T. and {Carballo}, R. and {Carlucci}, T. and {Carnerero}, M.~I. and {Carrasco}, J.~M. and {Casamiquela}, L. and {Castellani}, M. and {Castro-Ginard}, A. and {Chaoul}, L. and {Charlot}, P. and {Chemin}, L. and {Chiaramida}, V. and {Chiavassa}, A. and {Chornay}, N. and {Comoretto}, G. and {Contursi}, G. and {Cooper}, W.~J. and {Cornez}, T. and {Cowell}, S. and {Crifo}, F. and {Cropper}, M. and {Crosta}, M. and {Crowley}, C. and {Dafonte}, C. and {Dapergolas}, A. and {David}, M. and {David}, P. and {de Laverny}, P. and {De Luise}, F. and {De March}, R. and {De Ridder}, J. and {de Souza}, R. and {de Torres}, A. and {del Peloso}, E.~F. and {del Pozo}, E. and {Delbo}, M. and {Delgado}, A. and {Delisle}, J. -B. and {Demouchy}, C. and {Dharmawardena}, T.~E. and {Di Matteo}, P. and {Diakite}, S. and {Diener}, C. and {Distefano}, E. and {Dolding}, C. and {Edvardsson}, B. and {Enke}, H. and {Fabre}, C. and {Fabrizio}, M. and {Faigler}, S. and {Fedorets}, G. and {Fernique}, P. and {Fienga}, A. and {Figueras}, F. and {Fournier}, Y. and {Fouron}, C. and {Fragkoudi}, F. and {Gai}, M. and {Garcia-Gutierrez}, A. and {Garcia-Reinaldos}, M. and {Garc{\'\i}a-Torres}, M. and {Garofalo}, A. and {Gavel}, A. and {Gavras}, P. and {Gerlach}, E. and {Geyer}, R. and {Giacobbe}, P. and {Gilmore}, G. and {Girona}, S. and {Giuffrida}, G. and {Gomel}, R. and {Gomez}, A. and {Gonz{\'a}lez-N{\'u}{\~n}ez}, J. and {Gonz{\'a}lez-Santamar{\'\i}a}, I. and {Gonz{\'a}lez-Vidal}, J.~J. and {Granvik}, M. and {Guillout}, P. and {Guiraud}, J. and {Guti{\'e}rrez-S{\'a}nchez}, R. and {Guy}, L.~P. and {Hatzidimitriou}, D. and {Hauser}, M. and {Haywood}, M. and {Helmer}, A. and {Helmi}, A. and {Sarmiento}, M.~H. and {Hidalgo}, S.~L. and {Hilger}, T. and {H{\l}adczuk}, N. and {Hobbs}, D. and {Holland}, G. and {Huckle}, H.~E. and {Jardine}, K. and {Jasniewicz}, G. and {Jean-Antoine Piccolo}, A. and {Jim{\'e}nez-Arranz}, {\'O}. and {Jorissen}, A. and {Juaristi Campillo}, J. and {Julbe}, F. and {Karbevska}, L. and {Kervella}, P. and {Khanna}, S. and {Kontizas}, M. and {Kordopatis}, G. and {Korn}, A.~J. and {K{\'o}sp{\'a}l}, {\'A}. and {Kostrzewa-Rutkowska}, Z. and {Kruszy{\'n}ska}, K. and {Kun}, M. and {Laizeau}, P. and {Lambert}, S. and {Lanza}, A.~F. and {Lasne}, Y. and {Le Campion}, J. -F. and {Lebreton}, Y. and {Lebzelter}, T. and {Leccia}, S. and {Leclerc}, N. and {Lecoeur-Taibi}, I. and {Liao}, S. and {Licata}, E.~L. and {Lindstr{\o}m}, H.~E.~P. and {Lister}, T.~A. and {Livanou}, E. and {Lobel}, A. and {Lorca}, A. and {Loup}, C. and {Madrero Pardo}, P. and {Magdaleno Romeo}, A. and {Managau}, S. and {Mann}, R.~G. and {Manteiga}, M. and {Marchant}, J.~M. and {Marconi}, M. and {Marcos}, J. and {Marcos Santos}, M.~M.~S. and {Mar{\'\i}n Pina}, D. and {Marinoni}, S. and {Marocco}, F. and {Marshall}, D.~J. and {Martin Polo}, L. and {Mart{\'\i}n-Fleitas}, J.~M. and {Marton}, G. and {Mary}, N. and {Masip}, A. and {Massari}, D. and {Mastrobuono-Battisti}, A. and {Mazeh}, T. and {McMillan}, P.~J. and {Messina}, S. and {Michalik}, D. and {Millar}, N.~R. and {Mints}, A. and {Molina}, D. and {Molinaro}, R. and {Moln{\'a}r}, L. and {Monari}, G. and {Mongui{\'o}}, M. and {Montegriffo}, P. and {Montero}, A. and {Mor}, R. and {Mora}, A. and {Morbidelli}, R. and {Morel}, T. and {Morris}, D. and {Muraveva}, T. and {Murphy}, C.~P. and {Musella}, I. and {Nagy}, Z. and {Noval}, L. and {Oca{\~n}a}, F. and {Ogden}, A. and {Ordenovic}, C. and {Osinde}, J.~O. and {Pagani}, C. and {Pagano}, I. and {Palaversa}, L. and {Palicio}, P.~A. and {Pallas-Quintela}, L. and {Panahi}, A. and {Payne-Wardenaar}, S. and {Pe{\~n}alosa Esteller}, X. and {Penttil{\"a}}, A. and {Pichon}, B. and {Piersimoni}, A.~M. and {Pineau}, F. -X. and {Plachy}, E. and {Plum}, G. and {Poggio}, E. and {Pr{\v{s}}a}, A. and {Pulone}, L. and {Racero}, E. and {Ragaini}, S. and {Rainer}, M. and {Raiteri}, C.~M. and {Rambaux}, N. and {Ramos}, P. and {Ramos-Lerate}, M. and {Re Fiorentin}, P. and {Regibo}, S. and {Richards}, P.~J. and {Rios Diaz}, C. and {Ripepi}, V. and {Riva}, A. and {Rix}, H. -W. and {Rixon}, G. and {Robichon}, N. and {Robin}, A.~C. and {Robin}, C. and {Roelens}, M. and {Rogues}, H.~R.~O. and {Rohrbasser}, L. and {Romero-G{\'o}mez}, M. and {Rowell}, N. and {Royer}, F. and {Ruz Mieres}, D. and {Rybicki}, K.~A. and {Sadowski}, G. and {S{\'a}ez N{\'u}{\~n}ez}, A. and {Sagrist{\`a} Sell{\'e}s}, A. and {Sahlmann}, J. and {Salguero}, E. and {Samaras}, N. and {Sanchez Gimenez}, V. and {Sanna}, N. and {Santove{\~n}a}, R. and {Sarasso}, M. and {Schultheis}, M. and {Sciacca}, E. and {Segol}, M. and {Segovia}, J.~C. and {S{\'e}gransan}, D. and {Semeux}, D. and {Shahaf}, S. and {Siddiqui}, H.~I. and {Siebert}, A. and {Siltala}, L. and {Silvelo}, A. and {Slezak}, E. and {Slezak}, I. and {Smart}, R.~L. and {Snaith}, O.~N. and {Solano}, E. and {Solitro}, F. and {Souami}, D. and {Souchay}, J. and {Spagna}, A. and {Spina}, L. and {Spoto}, F. and {Steele}, I.~A. and {Steidelm{\"u}ller}, H. and {Stephenson}, C.~A. and {S{\"u}veges}, M. and {Surdej}, J. and {Szabados}, L. and {Szegedi-Elek}, E. and {Taris}, F. and {Taylor}, M.~B. and {Teixeira}, R. and {Tolomei}, L. and {Tonello}, N. and {Torra}, F. and {Torra}, J. and {Torralba Elipe}, G. and {Trabucchi}, M. and {Tsounis}, A.~T. and {Turon}, C. and {Ulla}, A. and {Unger}, N. and {Vaillant}, M.~V. and {van Dillen}, E. and {van Reeven}, W. and {Vanel}, O. and {Vecchiato}, A. and {Viala}, Y. and {Vicente}, D. and {Voutsinas}, S. and {Weiler}, M. and {Wevers}, T. and {Wyrzykowski}, {\L}. and {Yoldas}, A. and {Yvard}, P. and {Zhao}, H. and {Zorec}, J. and {Zucker}, S. and {Zwitter}, T.},
        title = "{Gaia Data Release 3. Summary of the content and survey properties}",
      journal = {\aap},
         year = 2023,
        month = jun,
       volume = {674},
          eid = {A1},
        pages = {A1},
          doi = {10.1051/0004-6361/202243940},
archivePrefix = {arXiv},
       eprint = {2208.00211},
 primaryClass = {astro-ph.GA},
       adsurl = {https://ui.adsabs.harvard.edu/abs/2023A&A...674A...1G}
}

@inproceedings{Akitaya2014honir,
  title = {HONIR: an optical and near-infrared simultaneous imager, spectrograph, and polarimeter for the 1.5-m Kanata telescope},
  author = {Akitaya, Hiroshi and Moritani, Yuki and Ui, Takahiro and Urano, Takeshi and Ohashi, Yuma and Kawabata, Koji S and Nakashima, Asami and Sasada, Mahito and Sakimoto, Kiyoshi and Harao, Tatsuya and others},
  booktitle = {Ground-based and Airborne Instrumentation for Astronomy V},
  volume = {9147},
  pages = {1520--1534},
  year = {2014},
  organization = {SPIE},
  address   = {Bellingham, WA},
  publisher = {SPIE}
}

@ARTICLE{2MASS,
       author = {{Skrutskie}, M.~F. and {Cutri}, R.~M. and {Stiening}, R. and {Weinberg}, M.~D. and {Schneider}, S. and {Carpenter}, J.~M. and {Beichman}, C. and {Capps}, R. and {Chester}, T. and {Elias}, J. and {Huchra}, J. and {Liebert}, J. and {Lonsdale}, C. and {Monet}, D.~G. and {Price}, S. and {Seitzer}, P. and {Jarrett}, T. and {Kirkpatrick}, J.~D. and {Gizis}, J.~E. and {Howard}, E. and {Evans}, T. and {Fowler}, J. and {Fullmer}, L. and {Hurt}, R. and {Light}, R. and {Kopan}, E.~L. and {Marsh}, K.~A. and {McCallon}, H.~L. and {Tam}, R. and {Van Dyk}, S. and {Wheelock}, S.},
        title = "{The Two Micron All Sky Survey (2MASS)}",
      journal = {\aj},
         year = 2006,
        month = feb,
       volume = {131},
       number = {2},
        pages = {1163-1183},
          doi = {10.1086/498708},
       adsurl = {https://ui.adsabs.harvard.edu/abs/2006AJ....131.1163S}
}

@ARTICLE{Pan-STARRS,
       author = {{Chambers}, K.~C. and {Magnier}, E.~A. and {Metcalfe}, N. and {Flewelling}, H.~A. and {Huber}, M.~E. and {Waters}, C.~Z. and {Denneau}, L. and {Draper}, P.~W. and {Farrow}, D. and {Finkbeiner}, D.~P. and {Holmberg}, C. and {Koppenhoefer}, J. and {Price}, P.~A. and {Rest}, A. and {Saglia}, R.~P. and {Schlafly}, E.~F. and {Smartt}, S.~J. and {Sweeney}, W. and {Wainscoat}, R.~J. and {Burgett}, W.~S. and {Chastel}, S. and {Grav}, T. and {Heasley}, J.~N. and {Hodapp}, K.~W. and {Jedicke}, R. and {Kaiser}, N. and {Kudritzki}, R. -P. and {Luppino}, G.~A. and {Lupton}, R.~H. and {Monet}, D.~G. and {Morgan}, J.~S. and {Onaka}, P.~M. and {Shiao}, B. and {Stubbs}, C.~W. and {Tonry}, J.~L. and {White}, R. and {Ba{\~n}ados}, E. and {Bell}, E.~F. and {Bender}, R. and {Bernard}, E.~J. and {Boegner}, M. and {Boffi}, F. and {Botticella}, M.~T. and {Calamida}, A. and {Casertano}, S. and {Chen}, W. -P. and {Chen}, X. and {Cole}, S. and {Deacon}, N. and {Frenk}, C. and {Fitzsimmons}, A. and {Gezari}, S. and {Gibbs}, V. and {Goessl}, C. and {Goggia}, T. and {Gourgue}, R. and {Goldman}, B. and {Grant}, P. and {Grebel}, E.~K. and {Hambly}, N.~C. and {Hasinger}, G. and {Heavens}, A.~F. and {Heckman}, T.~M. and {Henderson}, R. and {Henning}, T. and {Holman}, M. and {Hopp}, U. and {Ip}, W. -H. and {Isani}, S. and {Jackson}, M. and {Keyes}, C.~D. and {Koekemoer}, A.~M. and {Kotak}, R. and {Le}, D. and {Liska}, D. and {Long}, K.~S. and {Lucey}, J.~R. and {Liu}, M. and {Martin}, N.~F. and {Masci}, G. and {McLean}, B. and {Mindel}, E. and {Misra}, P. and {Morganson}, E. and {Murphy}, D.~N.~A. and {Obaika}, A. and {Narayan}, G. and {Nieto-Santisteban}, M.~A. and {Norberg}, P. and {Peacock}, J.~A. and {Pier}, E.~A. and {Postman}, M. and {Primak}, N. and {Rae}, C. and {Rai}, A. and {Riess}, A. and {Riffeser}, A. and {Rix}, H.~W. and {R{\"o}ser}, S. and {Russel}, R. and {Rutz}, L. and {Schilbach}, E. and {Schultz}, A.~S.~B. and {Scolnic}, D. and {Strolger}, L. and {Szalay}, A. and {Seitz}, S. and {Small}, E. and {Smith}, K.~W. and {Soderblom}, D.~R. and {Taylor}, P. and {Thomson}, R. and {Taylor}, A.~N. and {Thakar}, A.~R. and {Thiel}, J. and {Thilker}, D. and {Unger}, D. and {Urata}, Y. and {Valenti}, J. and {Wagner}, J. and {Walder}, T. and {Walter}, F. and {Watters}, S.~P. and {Werner}, S. and {Wood-Vasey}, W.~M. and {Wyse}, R.},
        title = "{The Pan-STARRS1 Surveys}",
      journal = {arXiv e-prints},
         year = 2016,
        month = dec,
          eid = {arXiv:1612.05560},
        pages = {arXiv:1612.05560},
          doi = {10.48550/arXiv.1612.05560},
archivePrefix = {arXiv},
       eprint = {1612.05560},
 primaryClass = {astro-ph.IM},
       adsurl = {https://ui.adsabs.harvard.edu/abs/2016arXiv161205560C}
}

@ARTICLE{Osaki1989,
       author = {{Osaki}, Yoji},
        title = "{A model for the superoutburst phenomenon of SU Ursae MAjoris stars.}",
      journal = {\pasj},
         year = 1989,
        month = jan,
       volume = {41},
        pages = {1005-1033},
       adsurl = {https://ui.adsabs.harvard.edu/abs/1989PASJ...41.1005O}
}

@ARTICLE{Hirose&Osaki1990,
       author = {{Hirose}, Masahito and {Osaki}, Yoji},
        title = "{Hydrodynamic Simulations of Accretion Disks in Cataclysmic Variables: Superhump Phenomenon in SU UMa Stars}",
      journal = {\pasj},
         year = 1990,
        month = feb,
       volume = {42},
        pages = {135-163},
       adsurl = {https://ui.adsabs.harvard.edu/abs/1990PASJ...42..135H}
}
\bibliographystyle{pasjtest1}

\end{document}